\documentstyle[12pt]{article}
\newcommand{\tr}{{\rm tr}}
\def\Sub#1{\vskip 1em \noindent {\bf#1}}
\newcommand{\BOX}{\hbox {$\sqcap$ \kern -1em $\sqcup$}}
\newcommand{\qed}{\hskip 3em \hbox{\BOX} \vskip 2ex}
\newcommand{\lgl}{\langle}
\newcommand{\rgl}{\rangle}
\newcommand{\dd}{\displaystyle}
\newcommand{\lds}{\ldots}
\newcommand{\mbk}{\medbreak}
\newcommand{\dta}{\delta}
\newcommand{\Dta}{\Delta}
\newcommand{\eps}{\epsilon}
\newcommand{\alp}{\alpha}

\begin{document}

\begin{center}
{\bf Vassiliev Invariants and the Loop States in Quantum Gravity\\ }
\vspace{0.5cm}
{Louis H.\ Kauffman\\ }
\vspace{0.3cm}
{Department of Mathematics, Statistics and Computer Science,\\
University of Illinois at Chicago, \\
Chicago, Illinois 60607-7045, USA\\
(email: U10451@uicvm.uic.edu)\\ }
\vspace{0.3 cm}
{to appear in {\sl Knots and Quantum Gravity},
Oxford U.\ Press}
\end{center}

\section{Introduction}

The purpose of this paper is to expose properties of Vassiliev
invariants by using the simplest of the approaches to the functional
integral definition of link invariants.  These methods are strong enough
to give the top  row evaluations of Vassiliev invariants for the
classical Lie algebras.  They give an insight into the structure of
these invariants without using the full perturbation expansion of the
integral.  One reason for examining the invariants in this light is the
possible applications to the loop variables approach to quantum gravity.
The same level of handling the functional integral is commonly used in
the loop transform for quantum gravity.

The paper is organized as follows.  Section 2 is a brief discussion of
the nature of the functional integral.  Section 3 details the
formalism of the functional integral that we shall use, and works out a
difference formula for changing crossings in the link invariant and a
formula for the change of framing.  These are applied to the case of
$SU(N)$ gauge group.  Section 4 defines the Vassiliev invariants and shows
how to formulate them in terms of the functional integral.  In
particular we derive the specific expressions for the ``top rows'' of
Vassiliev invariants corresponding to the fundamental representation of
$SU(N)$.  This gives a neat point of view on the results of Bar-Natan,
and also gives a picture of the structure of the graphical vertex
associated with the Vassiliev invariant.  We see that this vertex is not
just a transversal intersection of Wilson loops, but rather has the
structure of Casimir insertion (up to first order of approximation)
coming from the difference formula in the functional integral.  This
clarifies an issue raised by John Baez in \cite{[Baez]}.  Section
5 is a quick remark about the loop formalism for quantum gravity and its
relationships with the invariants studied in the previous sections.
This marks the beginning of a study that will be carried out in detail
elsewhere.

\section{Quantum mechanics and topology}

In \cite{[WIT89]} Edward Witten proposed a formulation of a class of
3-manifold
invariants as generalized Feynman integrals taking the form $Z(M)$ where
$$Z(M)=\int dA\exp\bigl[(ik/ 4\pi)S(M,A)\bigr].$$
Here $M$ denotes a 3-manifold without boundary and $A$ is a gauge field
(also called a qauge potential or gauge connection) defined on $M$.  The
gauge field is a one-form on $M$ with values in a representation of a
Lie algebra.  The group corresponding to this Lie algebra is said to be
the {\it gauge group\/} for this particular field.  In this integral the
``action'' $S(M,A)$ is taken to be the integral over $M$ of the trace
of the Chern-Simons three-form $CS=AdA+(2/3)AAA$.  (The product is the
wedge product of differential forms.)

Instead of integrating over paths, the integral $Z(M)$ integrates over
all gauge fields modulo gauge equivalence (see \cite{[AT79]}
for a discussion
of the definition and meaning of gauge equivalence.) This
generalization from paths to fields is characteristic of quantum field
theory.

Quantum field theory was designed in order to accomplish the
quantization of electromagnetism.  In quantum electrodynamics the
classical entity is the electromagnetic field.  The question posed in
this domain is to find the value of an amplitude for starting with one
field configuration and ending with another.
The analogue of all paths from point $a$ to point $b$ is ``all fields
from field $A$ to field $B$''.

Witten's integral $Z(M)$ is, in its form, a typical integral in quantum
field theory. In its content $Z(M)$ is highly unusual.  The formalism of
the integral, and its internal logic supports the existence of a large
class of topological invariants of 3-manifolds and associated invariants
of knots and links in these manifolds.

The invariants associated with this integral have been given rigorous
combinatorial descriptions (see
\cite{[RT91],[TW91],[KM91],[LI91],[KL93]}),
but questions and conjectures arising from the integral
formulation are still outstanding (see \cite{[GA93]}).

\section{Links and the Wilson loop}

We now look at the formalism of the Witten integral and see how it
implicates invariants of knots and links corresponding to each classical
Lie algebra.  We need the {\it Wilson loop.\/} The Wilson loop is an
exponentiated version of integrating the gauge field along a loop $K$ in
three space that we take to be an embedding (knot) or a curve with
transversal self-intersections.  For the purpose of this discussion, the
Wilson loop will be denoted by the notation $\big\lgl K|A\big\rgl$ to
denote the dependence on the loop $K$ and the field $A$.  It is usually
indicated by the symbolism tr$\biggl(P\exp\bigl(\dd \oint
KA\bigr)\biggr)$.

Thus $\big\lgl K|A\big\rgl=\tr\biggl(P\exp \bigl(\dd \oint
KA\bigr)\biggr)$.
Here the $P$ denotes path ordered integration.  The symbol $\tr$
denotes the trace of the resulting matrix.

With the help of the Wilson loop functional on knots and links, Witten
\cite{[WIT89]} writes down a functional integral for link invariants in a
3-manifold $M$:
\begin{eqnarray*}
Z(M,K) &=& \int dA\exp \bigl[(ik/4\pi)S(M,A)\bigr]
\tr\biggl(P\exp\bigl(\oint K A\bigr)\biggr)\\
&=& \int dA\exp\bigl[(ik/4\pi)S\bigr] \big\lgl K|A\big\rgl.\end{eqnarray*}
Here $S(M,A)$ is the Chern-Simons action, as in the previous
discussion.
We abbreviate $S(M,A)$ as $S$.
Unless otherwise mentioned, the manifold $M$ will be the
three-dimensional sphere $S^3$.

An analysis of the formalism of this functional integral reveals quite a
bit about its role in knot theory.  This analysis depends upon key facts
relating the curvature of the gauge field to both the Wilson loop and
the Chern-Simons Lagrangian.  To this end, let us recall the local
coordinate structure of the gauge field $A(x)$, where $x$ is a point in
three-space.  We can write $A(x)=A^a_k(x)T_adx^k$ where the index $a$
ranges from 1 to $m$ with the Lie algebra basis $\{T_1, T_2, T_3,\lds,
T_m\}$.  The index $k$ goes from 1 to 3.  For each choice of $a$ and
$k$, $A_a^k(x)$ is a smooth function defined on three-space.  In
$A(x)$ we sum over the values of repeated indices.  The Lie algebra
generators $T_a$ are actually matrices corresponding to a given
representation of an abstract Lie algebra.  We assume some properties of
these matrices as follows:
\begin{list}{}{}
\item
1. $[T_a, T_b]=if_{abc}T_c$ where $[x,y]=xy-yx$, and $f_{abc}$
(the matrix of structure constants) is totally antisymmetric.  There is
summation over repeated indices.
\mbk
\item
2. tr$(T_aT_b)=\dta_{ab}/2$ where $\dta_{ab}$ is the Kronecker
delta $(\dta_{ab}=1$ if $a=b$ and zero otherwise).
\end{list}
We also assume some facts about curvature.  (The reader may compare with
the exposition in \cite{[KA91]}.  But note the difference in conventions on
the use of $i$ in the Wilson loops and curvature definitions.)  The
first fact is the relation of Wilson loops and curvature for small
loops:

\Sub{Fact 1.}
The result of evaluating a Wilson loop about a very small
planar circle around a point $x$ is proportional to the area enclosed by
this circle times the corresponding value of the curvature tensor of the
gauge field evaluated at $x$.  The curvature tensor is written
$F^a_{rs}(x)T_adx^rdy^s$.  It is the local coordinate expression of
$dA+AA$.

\Sub{Application of Fact 1.} Consider a given Wilson loop
$\big\lgl K|A\big\rgl$.  Ask how its value will change if it is deformed
infinitesimally in the neighborhood of a point $x$ on the loop.
Approximate the change according to Fact 1, and regard the point $x$ as
the place of curvature evaluation.  Let $\dta\big\lgl K|A\big\rgl$
denote the change in the value of the loop.  $\dta\big\lgl K|A\big\rgl$
is given by the formula $\dta\big\lgl
K|A\big\rgl=dx^rdx^sF^a_{rs}(x)T_a\big\lgl K|A\big\rgl$.  This is the
first-order approximation to the change in the Wilson loop.

In this formula it is understood that the Lie algebra matrices $T_a$ are
to be inserted into the Wilson loop at the point $x$, and that we are
summing over repeated indices.  This means that each $T_a\big\lgl
K|A\big\rgl$ is a new Wilson loop obtained from the original loop
$\big\lgl K|A\big\rgl$ by leaving the form of the loop unchanged, but
inserting the matrix $T_a$ into the loop at the point $x$.  See the
figure below.

\vbox{\vskip 4 cm}  
$$\big\lgl K|A\big\rgl \qquad\qquad\qquad\qquad\qquad T_a\bigl\lgl
K|A\big\rgl$$

\Sub{Remark on Insertion.}
The Wilson loop is the limit, over partitions of the loop $K$, of
products of the matrices $\bigl(1+A(x)\bigr)$ where $x$ runs over the
partition.  Thus one can write symbolically,
$$\big\lgl K|A\big\rgl=\prod_{x\in K} \bigl(1+A(x)\bigr)=\prod_{x\in
K}\bigl(1+A^a_k(x)T_adx^k\bigr).$$
It is understood that a product of matrices around a closed loop
connotes the trace of the product.  The ordering is forced by the
one-dimensional nature of the loop.  Insertion of a given matrix into
this
product at a point on the loop is then a well-defined concept.  If $T$
is a given matrix then {\it it is understood that $T\big\lgl
K|A\big\rgl$ denotes the insertion of $T$ into some point of the
loop.\/} In the case above, it is understood from context of the formula
$ds^rdx^sF^a_{rs}(x)T_a\big\lgl K|A\big\rgl$ that the insertion is to be
performed at the point $x$ indicated in the argument of the curvature.

\Sub{Remark.} The previous remark implies the following formula for the
variation of the Wilson loop with respect to the gauge field:
\[  {\dta\bigl\lgl K|A\big\rgl\over\dta\bigl(A^a_k(x)\bigr)}
=dx^kT_a\big\lgl K|A\big\rgl.$$
Varying the Wilson loop with respect to the gauge field results in the
insertion of an infinitesimal Lie algebra element into the loop.

\Sub{Proof:} \begin{eqnarray*}
{\dta\big\lgl K|A\big\rgl\over\dta\bigl(A^a_k(x)\bigr)}
&=& {\dta\over\dta\bigl(A^a_k(x)\bigr)} \prod_{y\in K}
\bigl(1+A^a_k(y)T_ady^k\bigr)\\
&=&
\biggl[\,\prod_{y<x}\bigl(1+A^a_k(y)T_ady^k\bigr)\,\biggr][T_adx^k]
\biggl[\, \prod_{y>x}\bigl(1+A^a_k(y)T_ady^k\bigr)\,\biggr]\\
&=& dx^kT_a\big\lgl K|A\big\rgl.  \end{eqnarray*}
\hfill\qed

\Sub{Fact 2.} The variation of the Chern-Simons action $S$ with
respect to the gauge potential at a given point in three-space is
related to the values of the curvature tensor at that point by the
following
formula (where $\eps_{abc}$ is the epsilon symbol for three indices):
\[F^a_{rs}(x)=\eps_{rst}{\dta S\over\dta\bigl(A^a_t(x)\bigr)}.\]
With these facts at hand we are prepared to determine how the Witten
integral behaves under a small deformation of the loop $K$.

\Sub{Proposition 1.} (Compare \cite{[KA91]}.)
{\it All statements of equality in this proposition are up to order
$(1/k)^2$.\/}
\begin{list}{}{}
\item
1. Let $Z(K)=Z(K,S^3)$ and let $\dta Z(K)$ denote the change of
$Z(K)$ under an infinitesimal change in the loop $K$.  Then
$$\dta Z(K) =(4\pi i/k)\int dA
\exp\bigl[(ik/4\pi)S\bigr][\eps_{rst}dx^rdy^sdz^t]T_aT_a\big\lgl
K|A\big\rgl.$$
The sum is taken over repeated indices, and the insertion is taken of
the matrix products $T_aT_a$ at the chosen point $x$ on the loop $K$
that is regarded as the ``center'' of the deformation.  The volume
element $[\eps_{rst}dx^rdy^sdz^t]$ is taken with regard to the
infinitesimal directions of the loop deformation from this point on the
original loop.

\item
2. The same formula applies, with a different interpretation, to
the case where $x$ is a double point of transversal self intersection of
a loop $K$, and the deformation consists in shifting one of the crossing
segments perpendicularly to the plane of intersection so that the
self-intersection point disappears.  In this case, one $T_a$ is inserted
into each of the transversal crossing segments so that $T_aT_a\big\lgl
K|A\big\rgl$ denotes a Wilson loop with a self intersection at $x$ and
insertions of $T_a$ at $x+\eps_1$ and $x+\eps_2$, where $\eps_1$ and
$\eps_2$ denote small displacements along the two arcs of $K$ that
intersect at $x$.  In this case, the volume form is nonzero, with two
directions coming from the plane of movement of one arc, and the
perpendicular direction is the direction of the other arc.
\end{list}

\Sub{Proof:}
\begin{eqnarray*}  Z(K)&=& \int dA\exp\bigl[(ik/4\pi)S\bigr]
\dta\big\lgl K|A\big\rgl  \\
&=& \int dA\exp\bigl[(ik/4\pi)S\bigr]dx^rdy^sF^a_{rs}(x)T_a\big\lgl
K|A\big\rgl\qquad\rm{(Fact\; 1)}\\
&=& \int dA\exp\bigl[(ik/4\pi)S\bigr] dx^rdy^s\eps_{rst}
{\dta S\over \dta\bigl(A^a_t(x)\bigr)} T_a\big\lgl K|A\big\rgl
\qquad\rm{Fact\; 2)}  \\
&=& \int dA \bigl\{\exp\bigl[(ik/4\pi)S\bigr]{\dta
S\over \dta\bigl(A^a_t(x)\bigr)}\bigl\} \eps_{rst} dx^rdy^sT_a\big\lgl
K|A\big\rgl\\
&=& (-4\pi i/k)\int
dA{\dta\bigl\{\exp\bigl[(ik/4\pi)S\bigr]\bigr\}\over
\dta\bigl(A^a_t(x)\bigr)}
\eps_{rst}dx^rdy^sT_a\big\lgl K|A\big\rgl\\
&=& (4\pi i/k)\int dA\exp\bigl[(ik/4\pi)S\bigr]
\eps_{rst}dx^rdy^s{\dta\bigl\{T_a\big\lgl
K|A\big\rgl\bigr\}\over\dta\bigl(A^a_t(x)\bigr)}\\
&&\rm{(integration\; by \;parts)}\\
&=& (4\pi i/k)\int dA\exp \bigl[(ik/4\pi)S\bigr]
[\eps_{rst}dx^rdy^sdz^t]
T_aT_a\big\lgl K|A\big\rgl\\
&&\rm{(differentiating\; the \;Wilson\; loop).}\end{eqnarray*}
This completes the formalism of the proof.  In the case of part 2, the
change of interpretation occurs at the point in the argument when the
Wilson loop is differentiated.  Differentiating a self intersecting
Wilson loop at a point of self intersection is equivalent to
differentiating the corresponding product of matrices at a variable that
occurs at two points in the product (corresponding to the two places
where the loop passes through the point).  One of these derivatives
gives rise to a term with volume form equal to zero, the other term is
the one that is described in part 2.  This completes the proof of the
proposition.\hfill\qed

\Sub{Applying Proposition 1.}
As the formula of Proposition 1 shows, the integral $Z(K)$ is unchanged
if the movement of the loop does not involve three independent space
directions (since $\eps_{rst}dx^rdy^sdz^t$ computes a volume). This
means that $Z(K)=Z(S^3, K)$ is invariant under moves that slide the knot
along a plane.  In particular, this means that if the knot $K$ is given
in the nearly planar representation of a knot diagram, then $Z(K)$ is
invariant under regular isotopy of this diagram.  That is, it is
invariant under the Reidemeister moves II and III.  We expect more
complicated behavior under move I since this deformation does involve
three spatial directions.  This will be discussed momentarily.

We first determine the difference between $Z(K_+)$ and $Z(K_-)$ where
$K_+$
and $K_-$ denote the knots that differ only by switching a single
crossing.  We take the given crossing in $K_+$ to be the positive type,
and the crossing in $K_-$ to be of negative type.

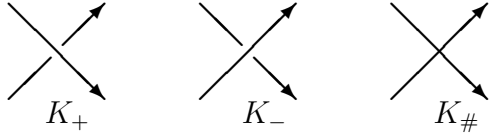
\begin{center}
\setlength{\unitlength}{0.00625in}%
\begin{picture}(400,100)(80,500)
\thicklines
\put( 80,600){\vector( 1,-1){ 80}}
\put(240,520){\vector( 1, 1){ 80}}
\put(285,555){\vector( 1,-1){ 35}}
\put(400,600){\vector( 1,-1){ 80}}
\put(400,520){\vector( 1, 1){ 80}}
\put(125,565){\vector( 1, 1){ 35}}
\put( 80,520){\line( 1, 1){ 35}}
\put(240,600){\line( 1,-1){ 35}}
\put(110,500){\makebox(0,0)[lb]{\raisebox{0pt}[0pt][0pt]{$K_+$}}}
\put(275,500){\makebox(0,0)[lb]{\raisebox{0pt}[0pt][0pt]{$K_-$}}}
\put(435,500){\makebox(0,0)[lb]{\raisebox{0pt}[0pt][0pt]{$K_\#$}}}
\end{picture}
\end{center} 
\noindent
The strategy for computing this difference is to use $K_\#$ as an
intermediate, where $K_\#$ {\it is the link with a transversal
self-crossing replacing the given crossing in $K_+$ or $K_-$\/}. Thus we
must consider $\Dta_+=Z(K_+)-Z(K_\#)$ and $\Dta_-=Z(K_-)-Z(K_\#)$.  The
second
part of Proposition 1 applies to each of these differences and gives
$$\Dta_+=(4\pi i/k)\int dA\exp\bigl[(ik/4\pi)S\bigr]
[\eps_{rst}dx^rdy^sdz^t] T_aT_a\big\lgl K_\#|A\big\rgl $$
where, by the description in Proposition 1, this evaluation is taken
along the loop $K_\#$ with the singularity and the $T_aT_a$ insertion
occurs along the two transversal arcs at the singular point.  The sign
of the volume element will be opposite for $\Dta_-$ and consequently we
have that
$$\Dta_+ + \Dta_- =0.$$
(The volume element $[\eps_{rst}dx^rdy^sdz^t]$ must be given a
conventional value in our calculations.  There is no reason to assign it
different absolute values for the cases of $\Dta_+$ and $\Dta_-$ since
they are symmetric except for the sign.)

Therefore $Z(K_+)-Z(K_\#)+\bigl(Z(K_-)-Z(K_\#)\bigr)=0$. Hence
$$Z(K_\#)=(1/2)\bigl(Z(K_+)+Z(K_-)\bigr).$$

\begin{center}
\setlength{\unitlength}{0.0125in}%
\begin{picture}(280,40)(60,680)
\thicklines
\put( 38,698){\makebox(0,0)[lb]{\raisebox{0pt}[0pt][0pt]{$Z($}}}
\put( 70,720){\vector( 1,-1){ 40}}
\put( 70,680){\vector( 1, 1){ 40}}
\put( 120,698){\makebox(0,0)[lb]
{\raisebox{0pt}[0pt][0pt]{$)\quad =\quad {1\over 2}\bigl(Z($}}}
\put(195,720){\vector( 1,-1){ 40}}
\put(220,705){\vector( 1, 1){ 15}}
\put(195,680){\line( 1, 1){ 15}}
\put( 240,698){\makebox(0,0)[lb]{\raisebox{0pt}[0pt][0pt]
{$\bigr)\; +\; Z\bigl( $}}}
\put(290,720){\line( 1,-1){ 15}}
\put(315,695){\vector( 1,-1){ 15}}
\put(290,680){\vector( 1, 1){ 40}}
\put( 335,698){\makebox(0,0)[lb]{\raisebox{0pt}[0pt][0pt]{$)\bigr)$}}}
\end{picture}
\end{center}
\noindent
This result is central to our further calculations.  It tells us that
the evaluation of a singular Wilson loop can be replaced with the
average of the results of resolving the singularity in the two possible
ways.

Now we are interested in the difference $Z(K_+)-Z(K_-)$:
$$Z(K_+)-Z(K_-) = \Dta_+ -\Dta_- = 2\Dta+ $$
$$ = (8\pi i/k)\int dA\exp \bigl[(ik/4\pi)S\bigr]
[\eps_{rst}dx^rdy^sdz^t]
T_a T_a\big\lgl K_\#|A\big\rgl.
$$

\Sub{Volume Convention.} It is useful to make a specific convention
about
the volume form.
We take
$$[\eps_{rst}dx^rdy^sdz^t]=\frac{1}{2} \ \rm{ for } \ \Dta_+ \ \rm{ and
}
\ -\frac{1}{2} \ \rm{ for } \ \Dta_-.$$
Thus
$$Z(K_+)-Z(K_-)=(4\pi i/k)\int dA\exp \bigl[(ik/4\pi)S\bigr]
T_aT_a\big\lgl K_\#|A\big\rgl.$$

\Sub{Integral Notation.} Let $Z(T_aT_aK_\#)$ denote the integral
$$Z(T_aT_aK_\#)= \int dA \exp\bigl[(ik/4\pi)S\bigr] T_aT_a\big\lgl
K_\#|A\big\rgl.$$

\Sub{Difference Formula.} Write the difference formula in abbreviated
form
$$Z(K_+)-Z(K_-)=(4\pi i/k) Z(T_aT_aK_\#).$$
This formula is the key to unwrapping many properties of the knot
invariants.  For diagrammatic work it is convenient to rewrite the
difference equation in the form shown below.  The crossings denote small
parts of otherwise identical larger diagrams, and the Casimir insertion
$T_aT_aK_\#$ is indicated with crossed lines entering a disk labelled
$C$.
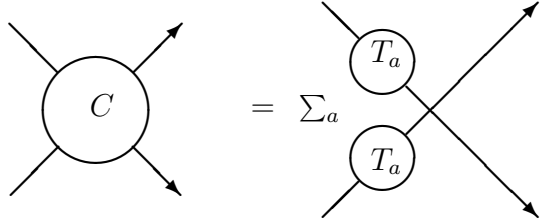
\begin{center}
\setlength{\unitlength}{0.00625in}%
\begin{picture}(440,160)(80,440)
\thicklines
\put(160,520){\circle{100}}
\put(400,560){\circle{50}}
\put(400,480){\circle{50}}
\put(192,552){\vector( 1, 1){ 40}}
\put(191,489){\vector( 1,-1){ 40}}
\put(128,552){\line(-1, 1){ 40}}
\put(129,489){\line(-1,-1){ 40}}
\put(420,540){\vector( 1,-1){110}}
\put(420,500){\vector( 1, 1){110}}
\put(380,580){\line(-1, 1){30}}
\put(380,460){\line(-1,-1){30}}
\put(155,515){\makebox(0,0)[lb]{\raisebox{0pt}[0pt][0pt]{$C$}}}
\put(290,515){\makebox(0,0)[lb]{\raisebox{0pt}[0pt][0pt]{$ =$}}}
\put(330,515){\makebox(0,0)[lb]{\raisebox{0pt}[0pt][0pt]{$\sum_a $}}}
\put(390,560){\makebox(0,0)[lb]{\raisebox{0pt}[0pt][0pt]{$T_a$}}}
\put(390,470){\makebox(0,0)[lb]{\raisebox{0pt}[0pt][0pt]{$T_a$}}}
\end{picture}
\end{center}

\Sub{The Casimir.}
The element $\sum_aT_aT_a$ of the universal enveloping algebra
is called the Casimir.  Its key
property is that it is in the center of the algebra.  Note that by our
conventions tr$(\sum_aT_aT_a)=\sum_a\Dta_{aa}/2=d/2$ where $d$ {\it is
the dimension of the Lie algebra.\/} This implies that an unknotted loop
with one singularity and a Casimir insertion will have $Z$-value $d/2$.

\vbox{\vskip 4 cm}  

\noindent
In fact, for the classical semi-simple Lie algebras {\it one can choose
a basis so that the Casimir is a diagonal matrix with identical values
$(d/2D)$ on its diagonal.\/} $D$ is the dimension of the representation.
We then have the general formula: $Z(T_aT_aK_\#^{\rm{loc}})=(d/2D) Z(K)$
for any knot $K$.  Here $K_\#^{\rm{loc}}$ denotes the singular knot
obtained by placing a local self-crossing loop in $K$ as shown below:

\vbox{\vskip 4 cm}  

\noindent
Note that $Z(K_\#^{\rm{loc}})=Z(K)$.  (Let the flat loop shrink to
nothing.  The Wilson loop is still defined on a loop with an isolated
cusp
and it is equal to the Wilson loop obtained by smoothing that cusp.)

Let $K_+^{\rm{loc}}$ denote the result of adding a positive local curl
to
the knot $K$, and $K_-^{\rm{loc}}$ the result of adding a negative local
curl to $K$.

\vbox{\vskip 4 cm}   

\noindent
Then by the above discussion and the difference formula, we have
\begin{eqnarray*}
Z(K_+^{\rm{loc}}) &= Z(K_\#^{\rm{loc}})+(2\pi i/k)
Z(T_aT_aK_\#^{\rm{loc}})\\
&= Z(K) +(2\pi i/k)(d/2D) Z(K).\end{eqnarray*}
Thus,
$$Z(K_+^{\rm{loc}})=\bigl(1+(\pi i/k)(d/D)\bigr) Z(K).$$
Similarly,
$$Z(K_-^{\rm{loc}})=\bigl(1-(\pi i/k)(d/D)\bigr) Z(K).$$

These calculations show how the difference equation, the Casimir, and
properties of Wilson loops determine the framing factors for the knot
invariants.  In some cases we can use special properties of the Casimir
to obtain skein relations for the knot invariant.

For example, in the fundamental representation of the Lie algebra for
$SU(N)$ the Casimir obeys the following equation (see
\cite{[KA91],[BN92]}):
$$\sum _a(T_a)_{ij}(T_a)_{kl}
=\biggl(\frac{1}{2}\biggr)\dta_{il}\dta_{jk}-\biggl(\frac{1}{2N}\biggr)
\dta_{ij}\dta_{kl}.$$
Hence
$$Z(T_aT_aK_\#)=\biggl(\frac{1}{2}\biggr)Z(K_0)-
\biggr(\frac{1}{2N}\biggr)Z(K_\#)$$
where $K_0$ denotes the result of smoothing a crossing as shown below:

\vbox{\vskip 4 cm}  

\noindent
Using $Z(K_\#)=\bigl(Z(K_+)+Z(K_-)\bigr)\big/2$ and the difference
identity, we obtain
$$Z(K_+)-Z(K_-) = (4\pi
i/k)\biggl\{\bigl(\frac{1}{2}\bigr)Z(K_0)-\biggl(\frac{1}{2N}\biggr)
\bigl[\bigl(Z(K_+)+Z(K_-)\bigr)\big/2\bigr]\biggr\}.$$
Hence
\[
(1+\pi i/Nk)Z(K_+) - (1-\pi i/Nk)Z(K_-) = (2\pi i/k)Z(K_o) \]
or
\[ e\biggl(\frac{1}{N}\biggr)Z(K_+) -
e\biggl(-\frac{1}{N}\biggr)Z(K_-)
= \bigl\{e(1)-e(-1)\bigr\}Z(K_0) \]
where $e(x)=\exp\bigl((\pi i/k)x\bigr)$ taken up
to $O(1/k^2)$.

Here $d=N^2-1$ and $D=N$, so the framing factor is
$$\alp
=\bigl(1+(\pi
i/k)\bigl((N^2-1)\big/N)\bigr)=e\bigl(
(N-(1/N)\bigr).$$
Therefore, if $P(K)=\alp^{-w(K)}Z(K)$ denotes the normalized invariant
of ambient isotopy associated with $Z(K)$ (with $w(K)$ the sum of the
crossing signs of $K$), then
$$\alp
e(1/N)P(K_+)-\alp^{-1}e(-1/N)P(K_-)=\bigl\{e(1)-e(-1)\bigr\}P(K_0).$$
Hence
$$e(N)P(K_+)-e(-N)P(K_-)=\bigl\{e(1)-e(-1)\bigr\}P(K_0).$$
This last equation shows that $P(K)$ is a specialization of the Homfly
polynomial for arbitrary $N$, and that for $N=2$ $\bigl(SU(2)\bigr)$ it
is a specialization of the Jones polynomial.

\section{Graph invariants and Vassiliev invariants}

We now apply this integral formalism to the structure of rigid vertex
graph invariants that arise naturally in the context of knot
polynomials.  If $V(K)$ is a (Laurent polynomial valued, or, more
generally,  commutative ring valued) invariant of knots, then it can be
naturally extended to an invariant of rigid vertex graphs by defining
the
invariant of graphs in terms of the knot invariant via an ``unfolding''
of the vertex as indicated below \cite{[KV92]}:
$$V(K_\$)=aV(K_+)+bV(K_-)+cV(K_0).$$

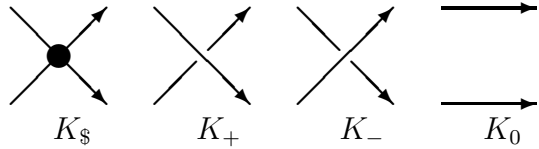
\begin{center}
\setlength{\unitlength}{0.00625in}%
\begin{picture}(480,100)(40,440)
\thicklines
\put( 120,500){\circle*{20}}
\put( 80,460){\vector( 1, 1){ 80}}
\put( 80,540){\vector( 1,-1){ 80}}
\put(200,540){\vector( 1,-1){ 80}}
\put(245,505){\vector( 1, 1){ 35}}
\put(320,460){\vector( 1, 1){ 80}}
\put(365,495){\vector( 1,-1){ 35}}
\put(440,540){\vector( 1, 0){ 80}}
\put(440,460){\vector( 1, 0){ 80}}
\put(320,540){\line( 1,-1){ 35}}
\put(200,460){\line( 1, 1){ 35}}
\put( 115,430){\makebox(0,0)[lb]{\raisebox{0pt}[0pt][0pt]{$K_\$ $}}}
\put(235,430){\makebox(0,0)[lb]{\raisebox{0pt}[0pt][0pt]{$K_+$}}}
\put(355,430){\makebox(0,0)[lb]{\raisebox{0pt}[0pt][0pt]{$K_-$}}}
\put(475,430){\makebox(0,0)[lb]{\raisebox{0pt}[0pt][0pt]{$K_0$}}}
\end{picture}   
\end{center}
\noindent
Here $K_\$$ indicates an embedding with a transversal 4-valent vertex
(\$).  We use the symbol \$ to distinguish this choice of vertex
designation from the previous one involving a self-crossing
Wilson loop.

Formally, this means that we define $V(G)$ for an embedded 4-valent
graph $G$ by taking the sum over $a^{i_+(S)}b^{i_-(S)}c^{i_0(S)}V(S)$
for
all knots $S$ obtained from $G$ by replacing a node of $G$ with either a
crossing of positive or negative type, or with a smoothing (denoted
$0$).  It is not hard to see that if $V(K)$ is an ambient isotopy
invariant of knots, then this extension is a rigid vertex isotopy
invariant of graphs.  In rigid vertex isotopy the cyclic order at the
vertex is preserved, so that the vertex behaves like a rigid disk with
flexible strings attached to it at specific points.

There is a rich class of graph invariants that cen be studied in this
manner.  The {\it Vassiliev invariants\/} \cite{[BN92],[BL91],[V90]}
constitute the important special case of these graph invariants where
$a=+1$, $b=-1$ and $c=0$.  Thus $V(G)$ is a Vassiliev invariant if
$$V(K_\$)=V(K_+)-V(K_-).$$
$V(G)$ is said to be the {\it finite type $k$ if $V(G)=0$ whenever\/}
$\#(G)>k$ where $\#(G)$ denotes the number of 4-valent nodes in the
graph $G$.  If $V$ is the finite type $k$, then $V(G)$ {\it is
independent of the embedding type of the graph $G$ when $G$ has exactly
$k$ nodes.\/} This follows at once from the definition of finite type.
The values of $V(G)$ on all the graphs of $k$ nodes is called the {\it
top row\/} of the invariant $V$.

For purposes of enumeration it is convenient to use chord diagrams to
enumerate and indicate the abstract graphs.  A chord diagram consists in
an oriented circle with an even number of points marked along it.  These
points are paired with the pairing indicated by arcs or chords
connecting the paired points.  See the figure below.

\vbox{\vskip 4 cm}   

\noindent
This figure illustrates the process of associating a chord diagram to a
given embedded 4-valent graph.  Each transversal self intersection in
the embedding is matched to a pair of points in the chord diagram.

\section{Vassiliev invariants from the functional integral}

In order to examine the Vassiliev invariants associated with the
functional integral, we must first normalize these invariants to
invariants of ambient isotopy, and then consider the structure of the
difference between positive and negative crossings.  The framed
difference formula provides the necessary information for obtaining the
top row.

We have shown that $Z(K_+^{\rm{loc}})=\alp Z(K)$ with $\alp=e(d/D)$.
Hence $$P(K)=\alp^{-w(K)}Z(K)$$
is an ambient isotopy invariant.  The
equation
\[ Z(K_+)-Z(K_-)=(4\pi i/k)Z(T_aT_aK_\#) \]
implies that if
$w(K_+)=w+1$, then we have the {\it ambient isotopy difference
formula:\/}
$$P(K_+)-P(K_-)=\alp^{-w}(4\pi
i/k)\bigl\{Z(T_aT_aK_\#)-(d/2D)Z(K_\#)\bigr\}.$$
We leave the proof of this formula as an exercise for the reader.

This formula tells us that for the Vassiliev invariant associated with
$P$ we have
$$P(K_\$)=\alp^{-w}(4\pi
i/k)\bigl\{Z(T_aT_aK_\#)-(d/2D)Z(K_\#)\bigr\}.$$
Furthermore, if $V_j(K)$ denotes the coefficient of $(4\pi i/k)^j$ in
the expansion of $P(K)$ in powers of $(1/k)$, then the ambient
difference formula implies that $(1/k)^j$ divides $P(G)$ when $G$ has
$j$ or more nodes.  Hence $V_j(G)=0$ if $G$ has more than $j$ nodes.
Therefore $V_j(K)$ is a Vassiliev invariant of finite type.  (This
result was proved by Birman and Lin \cite{[BL91]} by different methods
and
by Bar-Natan \cite{[BN92]} by methods equivalent to ours.)

The fascinating thing is that the ambient difference formula,
appropriately interpreted, actually tells us how to compute $V_k(G)$
when $G$ has $k$ nodes.  Under these circumstances each node undergoes a
Casimir insertion, and because the Wilson loop is being evaluated
abstractly, independent of the embedding, we insert nothing else into
the loop.  Thus we take the pairing structure associated with the graph
(the so-called chord diagram) and use it as a prescription for obtaining
a trace of a sum of products of Lie algebra elements with $T_a$ and
$T_a$ inserted for each pair or a simple crossover for the pair
multiplied by $(d/2D)$.  This yields the graphical evaluation implied by
the recursion
$$V(G_\$)=\bigl\{V(T_aT_aG_\#)-(d/2D)V(G_\#)\bigr\}.$$

\vbox{\vskip 4 cm}  

\noindent
At each stage in the process one node of $G$ disappears or it is
replaced by these insertions.  After $k$ steps we have a fully inserted
sum of abstract Wilson loops, each of which can be evaluated by taking
the indicated trace.  This result is equivalent to Bar-Natan's result,
but it is very interesting to see how it follows from a minimal approach
to the Witten integral.

In particular, it follows from Bar-Natan \cite{[BN92]} and Kontsevich
\cite{[KO92]} that the condition of topological invariance is
translated into the fact that the Lie bracket is represented as a
commutator {\it and\/} that it is closed with respect to the Lie
algebra.  Diagrammatically we have:

\vbox{\vskip 4 cm}  

\noindent
Since
$$  T_a T_b - T_b T_a = \sum_c if_{abc}T_c $$
we obtain

\vbox{\vskip 4 cm}   

\noindent
This relationship on chord diagrams is the seed of all the topology.  In
particular, it implies the basic 4-term relation,

\vbox{\vskip 4 cm}     

\noindent

\Sub{Proof:}

\vbox{\vskip 4 cm}      
{}\hfill\qed

The presence of this relation on chord diagrams for $V_i(G)$ with
$\#(G)=i$ is the basis for the existence of a corresponding Vassiliev
invariant.  There is not room here to go into more detail about this
matter, and so we bring this discussion to a close.  Nevertheless,
it must be mentioned that this brings us to the core of the main
question about Vassiliev invariants: Are there non-trivial Vassiliev
invariants of knots and links that cannot be constructed through
combinations of Lie algebraic invariants? There are many other open
questions in this arena, all circling this basic problem.

\section{Quantum gravity---loop states}

We now discuss the relationship of Wilson loops and quantum gravity that
is forged in the theory of Ashtekar, Rovelli and Smolin \cite{[ASH92]}.
In
this theory the metric is expressed in terms of a spin connection $A$,
and quantization involves considering wavefunctions $\psi(A)$.  Smolin
and Rovelli analyze the loop transform $\hat\psi(K)=\dd\int dA
\psi(A)\big\lgl K|A\big\rgl$ where $\big\lgl K|A\big\rgl$ denotes the
Wilson loop for the knot or singular embedding $K$.  Differential
operators on the wavefunction can be referred, via integration by parts,
to corresponding statements about the Wilson loop.  It turns out that
the condition that $\hat\psi(K)$ be a knot invariant (without framing
dependence) is equivalent to the so-called diffeomorphism constraint
\cite{[SM88]} for these wave functions.  In this way, knots and weaves
and
their topological invariants become a language for representing a
state of quantum gravity.

The main point that we wish to make in the relationship of the Vassiliev
invariants to these loop states is that the Vassiliev vertex

\vbox{\vskip 4 cm}   

\noindent
satisfying

\vbox{\vskip 4 cm}   

\noindent
is not simply a transverse intersection of Wilson loops.  We have seen
that it follows from the difference formula that the Vassiliev vertex is
much more complex than this---that it involves the Casimir insertion at
the transverse intersection up to the first order of approximation, and
that this structure can be used to compute the top row of the
corresponding Vassiliev invariant.  This situation suggests that one
should amalgamate the formalism of the Vassiliev invariants with the
structure of the Poisson algebras of loops and insertions used in the
quantum gravity theory \cite{[SM88],[PUL93]}.  It also suggests taking
the formalism of these invariants (in the functional integral form)
quite seriously even in the absence of an appropriate measure theory.

One can begin to work backwards, taking the position that invariants
that do not ostensibly satisfy the diffeomorphism constraint (due to
change of value under framing change) nevertheless still define states
of a quantum gravity theory that is a modification of the
Ashtekar formulation.  This theory can be investigated by working the
transform methods backwards---from knots and links to differential
operators and differential geometry.

All these remarks are the seeds for another paper.  We close here, and
ask the reader to stay tuned for further developments.

\section*{Acknowledgements}
It gives the author pleasure to acknowledge the support of NSF Grant
Number DMS 9205277 and the Program for Mathematics and Molecular Biology
of the University of California at Berkeley, Berkeley, CA.

\end{document}